\documentclass[twoside]{dis08}
\usepackage[latin1]{inputenc}
\usepackage[dvips]{graphicx,epsfig,color}
\usepackage{wrapfig,rotating}
\usepackage{amssymb,amsmath,array}

\newcommand{\xp}{x_{I\!\!P}} 

\begin{document}
\title{Scaling properties in deep inelastic scattering}

\author{Christophe Royon$^{1}$, Guillaume Beuf$^{2}$, Robi Peschanski$^{2}$ and 
David Salek$^{3}$
%
%
\vspace{.3cm}\\
%
%
\vspace{.1cm}\\
1- CEA/IRFU/Service de physique des particules,  91191 
Gif-sur-Yvette cedex, France
\vspace{.1cm}\\
2- CEA/Service de physique th\'eorique,  91191 
Gif-sur-Yvette cedex, France
\vspace{.1cm}\\
3- Institute of Particle and Nuclear Physics, Charles University, Prague, 
Czech Republic
}

\maketitle

\begin{abstract}
Using the ``Quality Factor'' (QF) method, we analyse the scaling properties
of deep-inelastic processes
at HERA and fixed target experiments for $x\!\le \!10^{-2}.$
\end{abstract}

Geometric scaling~\cite{Stasto:2000er} is a remarkable empirical 
property 
verified by data on high energy deep inelastic scattering (DIS). 
One can  represent with 
reasonable 
accuracy the cross section $\sigma^{\gamma^*p}$ by the formula
$\sigma^{\gamma^*p}(Y,Q)=\sigma^{\gamma^*}(\tau)\  ,$
where $Q$ is the virtuality of the photon,
$Y$ the total rapidity in the ${\gamma^*}$-proton system and 
$\tau = \log Q^2-\log Q_s(Y) =  \log Q^2-\lambda 
Y\ $ 
is the scaling variable. In this paper, we will study different forms of
scalings predicted by theory and compare them to the data~\cite{us}.

\begin{footnotesize}
\begin{table} 
\begin{center}
\begin{tabular}{|c||c||c|} \hline
scaling& $\partial_L T$ & $L \partial T/ \partial Y$  \\
\hline\hline
$T(L-\lambda \sqrt{Y})$ & $\frac{\partial T}{\partial L}(L -\lambda
\sqrt{Y})=\frac{\partial T}{\partial L}
$  & $L \frac{\partial T}{\partial Y}= \frac{\lambda L}
{2 \sqrt{Y}} T$ \\
 & SCALING & APPR. SCALING \\ \hline
$T(L-\lambda Y/L$) & $\frac{\partial T}{\partial L}
(L-\lambda Y/L) = \left( 1 +
\frac{\lambda Y}{L^2} \right) \frac{\partial T}{\partial L}$ & $L \frac{\partial T}{\partial Y}(L - \lambda
Y/L)= -\lambda \frac{\partial T}{\partial Y}$
 \\
  & APPR. SCALING & SCALING \\ \hline
\end{tabular}
\caption{Approximate scalings in the case of the BK equation (RCI and RCII).}
\end{center}
\end{table}
\end{footnotesize}

\section{Scaling variables}
The stochastic extension of the Balitsky-Kovchegov equation~\cite{balitsky} for
dipole amplitude $T$ reads
\begin{eqnarray}
\frac{\partial T}{\partial Y} = \alpha_S \left[ \chi (-\partial_L)T -
T^2 +
\sqrt{\alpha_S^2 \kappa T}~ \nu(L,Y) \right]
\label{master}
\end{eqnarray}
where $\chi$ is the BFKL~\cite{bfkl} kernel, $L=\log Q^2$ and $\nu$  a
gaussian ``noise" corresponding to the fluctuation of the number of gluons in
the proton. The second term of Eq.~\ref{master} corresponds to the gluon
recombination and 
the last term to pomeron loops. We notice that we
obtain the BFKL LL equation when $\alpha_S$ is taken to be constant, the 
term in $T^2$ is neglected and $\kappa=0$ and the
Balitsky-Kovchegov equation~\cite{balitsky} when $\alpha_S$ is constant and $\kappa=0$.
Let us consider different cases.

When $\alpha_S$ is constant and $\kappa=0$, it is possible to show that the
solution of the Balitsky-Kovchegov equation does not depend independently on
$Y$ and $\log Q^2$ but on a combination of both, $\tau = L - \lambda Y$. This is
called ``fixed coupling" (FC) in the following.

When $\alpha_S$ is running ($\alpha_S \sim 1/\log Q^2$),
it is impossible that both $L \partial T/\partial Y$ and $\chi(-\partial_L)T$ follow
the same scaling at the same time. Two approximate solutions are found either in
$T(L-\lambda \sqrt{Y})$ or in $T(L-\lambda Y/L)$. In Table 1, we see that both
scalings cannot be satisfied at the same time and scaling is only exact 
when $L \sim \sqrt Y$. In the following, we call the two scalings $\tau = L - \lambda \sqrt{Y}$ and
$\tau = L - \lambda Y/L$, ``Running coupling
I" (RCI) and ``Running coupling II" (RCII) respectively. An extension of ``Running coupling
II" (called RCIIbis) was also considered by adding the additional parameters $\Lambda$ and $Y_0$
in $Q/\Lambda$ and $Y-Y_0$ instead of $Q$ and $Y$ respectively.
The third case is when $\alpha_S$ is constant, and we introduce the term in
$\sqrt{T}$. In that case, it can be shown that the scaling is 
\begin{wrapfigure}{r}{0.5\columnwidth}
\centerline{\includegraphics[width=0.4\columnwidth]{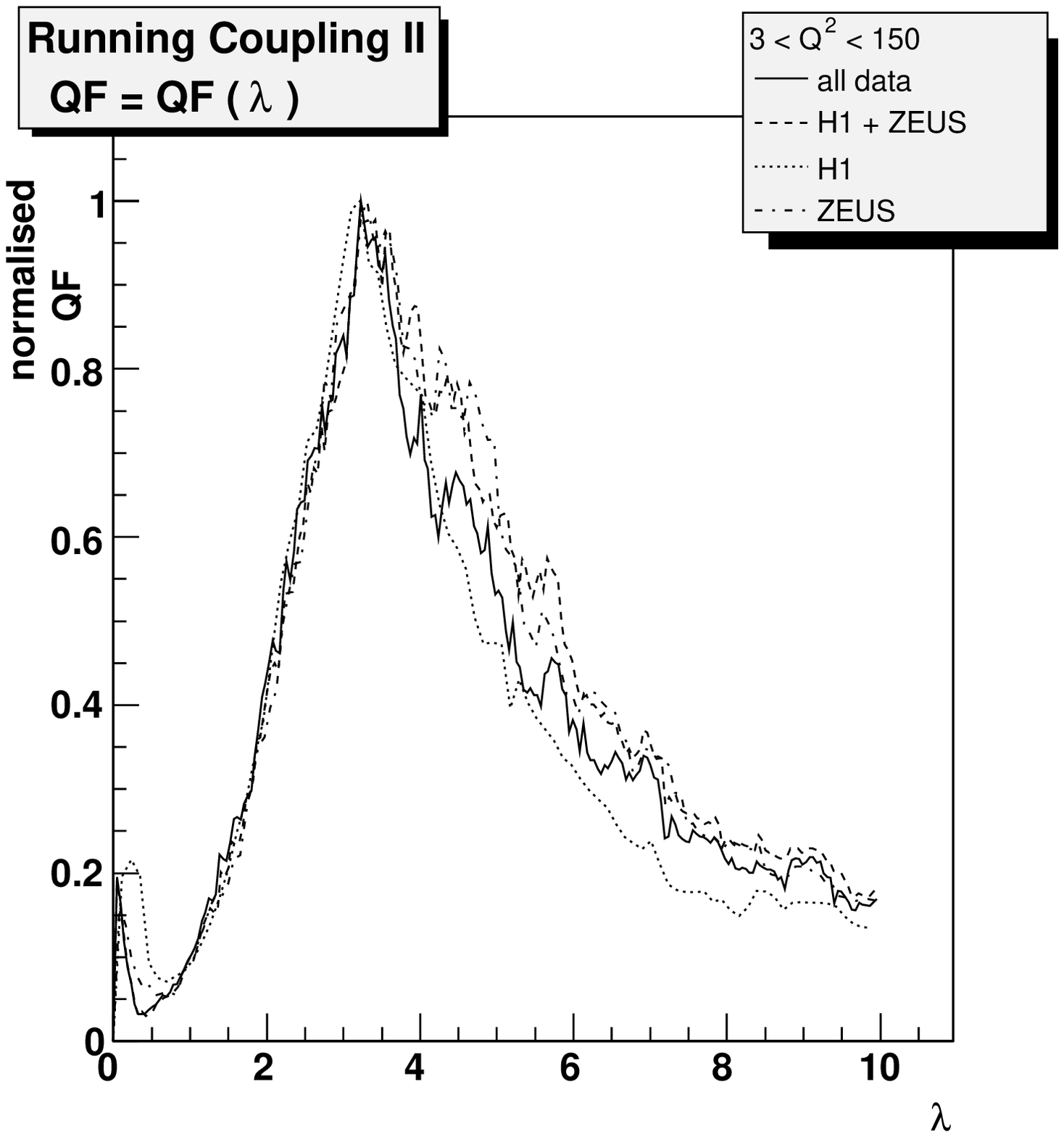}}
\centerline{\includegraphics[width=0.4\columnwidth]{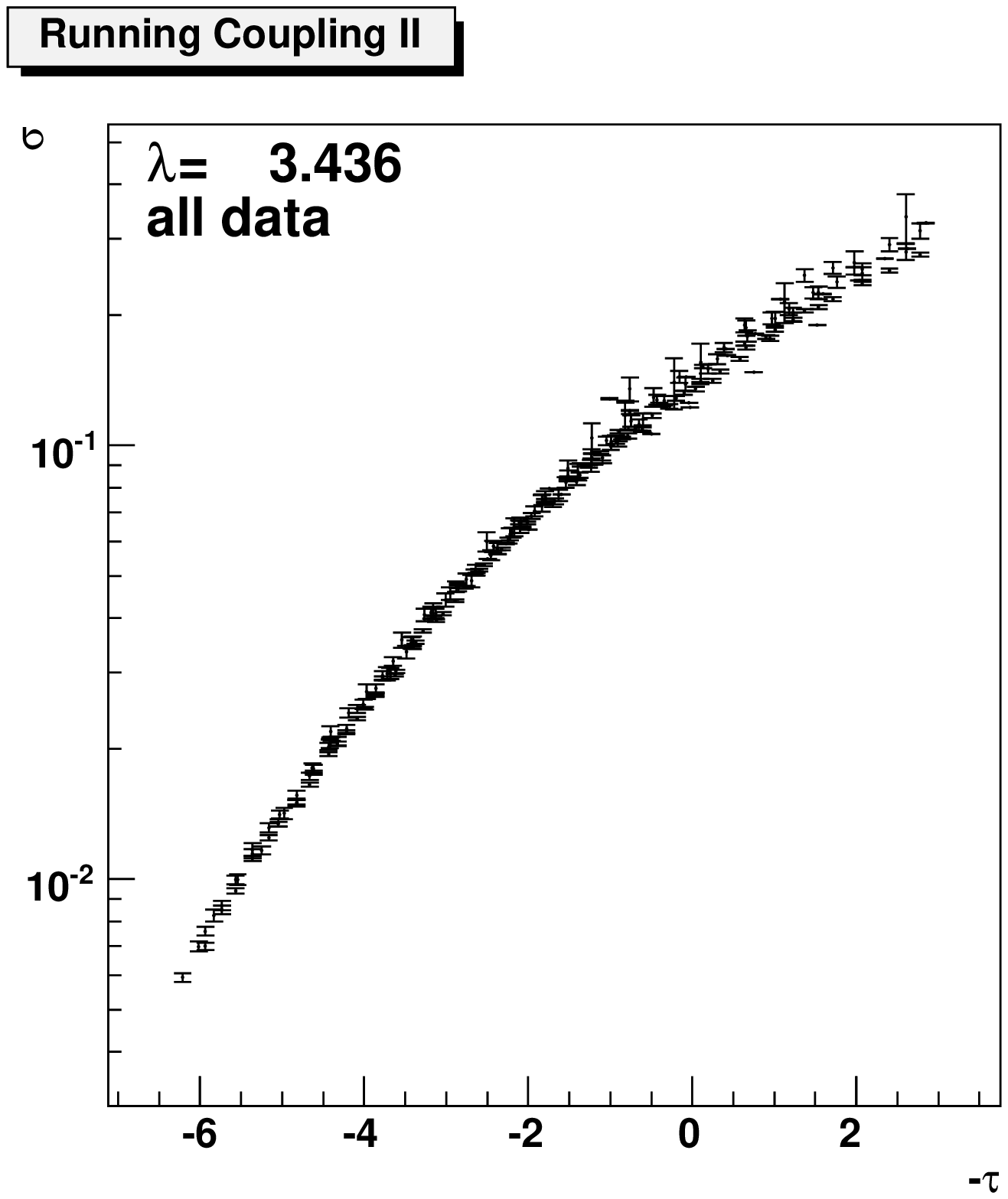}}
\caption{Normalised QF to 1. as a function of $\lambda$ and scaling curve 
with $\lambda$ fixed to the value corresponding to the best QF
for RCII ($\lambda=3.44$). A $Q^2>3$ GeV$^2$ cut was applied to the data.}
\label{rc2}
\end{wrapfigure}
$(L - \lambda Y)/\sqrt{Y}$, called diffusive scaling (DS) in the following.

\section{Scaling in DIS data}

\subsection{Quality factor}

The question rises how to quantify the quality of the different scalings and
compare them using the DIS data. The difficulty comes
from the fact that the $\tau$ dependence is not known and an estimator is needed
to know whether data points depend only on $\tau$ or not. This is why the concept of
quality factor (QF) was introduced~\cite{qf}.

The first step is the normalise the data sets ($F_2$ measurements for instance)
$v_i=\log(\sigma_i)$ and the scalings $u_i=\tau_i(\lambda)$ between 0 and 1
\footnote{Note that we take the logarithm of the cross section since it varies
by orders of magnitude.}. The $u_i$ are ordered. We define the
QF as
\begin{eqnarray}
QF(\lambda) = \left[ \sum_{i} \frac{(v_i-v_{i-1})^2 }
{(u_i-u_{i-1})^2+\epsilon^2} \right]^{-1}
\end{eqnarray}
where $\epsilon$ is a small term (taken as 0.0001) needed in the case when 
two data points have the same scaling (same $x$ and $Q^2$).
The method is then simple: we fit $\lambda$ to maximise QF to obtain
the best scaling.

\subsection{Scaling tests in DIS using $F_2$}
To test the different scalings, we first consider the proton structure function
measurements $F_2$ from H1, ZEUS, NMC and E665~\cite{H1}. We apply some cuts
on data $1 \le Q^2 \le 150$ GeV$^2$, $x \le 10^{-2}$ to stay in the
perturbative domain and also
avoid the region where
valence quark dominates, which leads to 217 data points. As an example, the
results for RCII are given in Fig.~\ref{rc2}. The different
scalings show quite good values of QF, while RCII is favoured and DS disfavoured
as it is indicated in Table 2.

\begin{table} 
\begin{center}
\begin{tabular}{|c||c|c|} \hline
scaling& Parameter & QF \\
\hline\hline
FC & $\lambda=-0.33$ & 1.63 \\ 
RCI & $\lambda=1.81$ & 1.62 \\ 
RCII & $\lambda=3.44$ & 1.69 \\ 
RCII bis & $\lambda=3.90$, $\Lambda = 0.30$, $Y_0 = -1.2$ & 1.82 \\
DS & $\lambda=0.36$ & 1.44 \\ \hline
\end{tabular}
\end{center}
\caption{QF for $F_2$ data ($Q^2>3$ GeV$^2$) and the different scalings.
The results are similar when taking all data with $Q^2>1$ GeV$^2$. }
\end{table}

\begin{wrapfigure}{r}{0.5\columnwidth}
\centerline{\includegraphics[width=0.4\columnwidth]{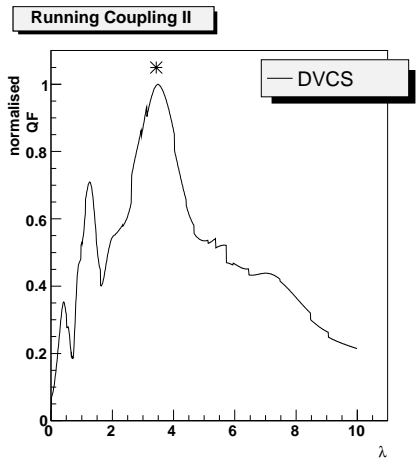}}
\caption{$\lambda$ dependence of the normalised QF
to 1. for DVCS data and for RCII. The star indicated the
values of $\lambda$ obtained with a fit to $F_2$, $Q^2>3$ GeV$^2$.
}\label{Fig2}
\end{wrapfigure}

To study the $Q^2$ dependence of the fitted parameters 
the data points are divided into four separate $Q^2$ samples:
$[1;3]$, $[3;10]$, $[10;35]$, and $[35;150]$~GeV$^2$. There is a slight increase
of the $\lambda$ parameter in the case of FC (from 0.28 to
0.40) while
RCI is quite flat at $\lambda \sim 1.8$. This can be easily understood
since RCI shows a natural $Q^2$ evolution. 
We notice a stronger increase in the case of
RCII ($\lambda$=1.6; 3.2; 3.5; 4.1 in the different
$Q^2$ regions) , showing the breaking of this scaling as a function
of $Q^2$. Since this scaling gives already the best QF, it would be worth to study the
breaking of scaling and introduce it in the model to further improve the description of the
data. 
The $\lambda$ parameter decreases strongly
(especially in the last $Q^2$ bin) for the DS from
0.46 to 0.11 which confirms the fact that this scaling leads to the worst
description of the data.

We also tested that the $F_2^c$ from charm data shows the same scalings
as for $F_2$. However the $F_2^c$ MRST and CTEQ parametrisations corresponding
to NLO QCD DGLAP fits do not show the same
scaling, which shows a different behavour between data and
parametrisation~\cite{us}.

\subsection{Fits to DVCS data from H1 and ZEUS}
After fitting all H1 and ZEUS $F_2$ data, it is worth studying whether the DVCS data measured by
the same experiments~\cite{DVCS} lead to the same results. The amount of data is smaller
(34 points for H1 and ZEUS requiring $x \le 0.01$ as for $F_2$ data) and the precision
on the $\lambda$ parameter will be weaker. The results of the fits can be found  in Fig.~2.
To facilitate the comparison between the results of the fits to $F_2$ and DVCS
data, a star is put  
at the position of the $\lambda$ value fitted to the H1+ZEUS $F_2$ data
with $Q^2$ in the range $[3;150]$~GeV$^2$. We note that 
the DVCS data lead to similar
$\lambda$ values to the $F_2$ data, showing the consistency of the 
scalings. 

\subsection{Implications for Diffraction and Vector Mesons}
In this section, we check if the scalings found in the previous section can also describe
diffractive and vector meson data. Since these data are much less precise than the $F_2$ or
DVCS data
and depend more on non-perturbative inputs (meson wave function, diffractive
parton distribution inputs...), we choose to impose the same values of parameters found in the previous section
and check if the scaling is also observed using this value. Concerning the
diffractive $F_2^D$ data~\cite{F2d}, we use $\beta d \sigma_{diff}^{\gamma^*
\rightarrow Xp}$, and the same definition of $\tau$, replacing $x$ by $\xp$,
$Q^2$ remaining the same. For vector meson data~\cite{vm}, we use the same scaling formulae as before replacing
$Q^2$ by $Q^2+M_V^2$ where $M_V$ is the mass of the vector meson. Both data sets
show the same scalings as for $F_2$, and the quality factors are similar~\cite{us}. As an
example, we give the scaling curves for $\rho$ data and the fixed coupling
scaling in Fig.~3.

As a phenomenological outlook, it seems useful to work out models dipole 
amplitude which could incorporate the successful scaling laws. On the theoretical ground, our phenomenological analysis can help to improve 
the theoretical analysis of scaling. 

\begin{wrapfigure}{r}{0.5\columnwidth}
\centerline{\includegraphics[width=0.4\columnwidth]{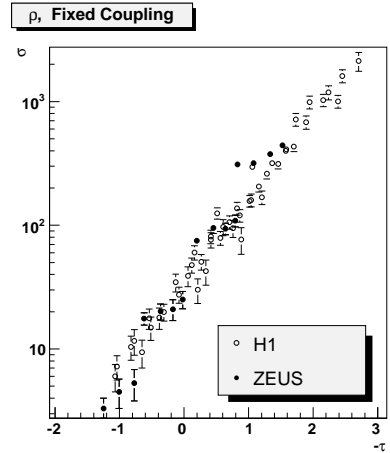}}
\caption{Scaling curves for
FC and $\rho$ data using the same $\lambda$ parameter as for the $F_2$ fit.} 
\end{wrapfigure}

\begin{footnotesize}


\end{footnotesize}



\begin{thebibliography}{99}
\bibitem{url} Slides: 
\verb$http://indico.cern.ch/contributionDisplay.$ \verb$py?contribId=184&sessionId=17&confId=24657$

\bibitem{Stasto:2000er}
A.~M. Sta\'sto, K.~Golec-Biernat, and J.~Kwiecinski,
Phys. Rev. Lett. {\bf 86}, 596 (2001);  K.~Golec-Biernat and M.~Wusthoff,
Phys.\ Rev.\ D {\bf 59}, 014017 (1999).
\bibitem{us}
G.~Beuf, R.~Peschanski, C.~Royon, D.~Salek, arXiv:0803.2186
and references therein;  G.~Beuf, arXiv:0803.2167.

\bibitem{balitsky}
I.~Balitsky,
Nucl. Phys. B {\bf 463}, 99 (1996);
Y.~V. Kovchegov,
Phys. Rev. D {\bf 61}, 074018 (2000).



\bibitem{bfkl} L. N. Lipatov, {\it Sov. J. Nucl. Phys.} {\bf 23} (1976) 338;
E. A. Kuraev, L. N. Lipatov and V. S. Fadin,
{\it Sov. Phys. JETP} {\bf 45} (1977) 199;
I. I. Balitsky and L. N. Lipatov,
{\it Sov. J. Nucl. Phys.} {\bf 28} (1978) 822. 

\bibitem{qf}
F.~Gelis, R.~Peschanski, G.~Soyez, L.~Schoeffel,
  Phys.\ Lett.\  B {\bf 647}, 376 (2007);  C.~Marquet, L.~Schoeffel,
  Phys.\ Lett.\  B {\bf 639}, 471 (2006).


\bibitem{H1}
C.~Adloff {\it et al.},
Eur.\ Phys.\ J.\ C {\bf 30} (2003) 1;
J.~Breitweg {\it et al.},
Phys.\ Lett.\ B {\bf 487} (2000) 273;
S.~Chekanov {\it et al.},
Phys.\ Rev.\ D {\bf 70} (2004) 052001;
M.~Arneodo {\it et al.},
Nucl.\ Phys.\ B {\bf 483} (1997) 3;
M.~R.~Adams {\it et al.},
Phys.\ Rev.\ D {\bf 54} (1996) 3006.
%

\bibitem{DVCS}
A.~Aktas {\it et al.}, Phys. Lett. B{\bf 659} (2008) 796-806;

\bibitem{F2d}
A.~Aktas {\it et al.}, Eur. Phys. J. C {\bf 48} (2006) 715-748;
S.~Chekanov {\it et al.}, Nucl. Phys. B {\bf 713} (2005) 3-80.

\bibitem{vm}
S.~Chekanov {\it et al.}, Nucl. Phys. B {\bf 718} (2005) 3-31;
A.~Aktas {\it et al.}, Eur. Phys. J. C {\bf 46} (2006) 585-603.









\end{thebibliography}
\end{document}